\documentclass[prb,aps,twocolumn,showpacs,preprintnumbers,amsmath,amssymb]{revtex4}
\usepackage{epsf}

\begin{document}
\draft
\newcommand{\ve}[1]{\boldsymbol{#1}}

\title{Structural relaxation and metal-insulator transition at the interface between
SrTiO$_3$ and LaAlO$_3$}
\author{Natalia Pavlenko$^{1,2}$ and Thilo Kopp$^1$}
\address{$^1$EKM, Universit\"at Augsburg, 86135 Augsburg, Germany \\
$^2$Institute for Condensed Matter Physics, 79011 Lviv, Ukraine}

\begin{abstract}
The electronic structure of interfaces between LaAlO$_3$ and SrTiO$_3$
is studied using local spin density approximation (LSDA) with intra-atomic
Coulomb repulsion (LSDA+U). We find that the nature of the interface metallic states 
is strongly affected 
by the type of the structure (sandwich or bilayer) and
by the termination surface of LaAlO$_3$. In all 
structures the atomic relaxation plays a crucial role in the electronic properties
of the system. While in sandwiches
the structural
relaxation produces a significant polarization in SrTiO$_3$ and
Jahn-Teller like splitting
of Ti $3d_{xy}$ orbitals, in AlO$_2$-terminated bilayers the relaxation occurs
primarily in LaAlO$_3$ and results in an insulator-metal transition which 
has been observed experimentally with
increasing thickness of the LaAlO$_3$ layer.
\end{abstract}


\date{\today}

\maketitle

\section{Introduction}

Interfaces in complex transition metal oxide heterostructures can play an important role 
for their physical properties \cite{dagotto}. This is especially important for heterostructures 
which contain nm-thick films deposited on structurally compatible substrate layers. 
The lattice mismatch and the charge polarity due to structural discontinuities at a film-substrate
interface are the driving forces of interfacial reconstruction which changes dramatically
the interface electronic state and possibly the physical behavior of the entire film.

The character of such a reconstruction can range from purely electronic to mixed electronic-lattice.
The electronic reconstruction occurs through a compensation of the interface polarity by 
charge self-doping. In addition, the lattice relaxation can modify 
the band structure (Jahn-Teller like effect)
and redistribute the charge between the electronic levels with consequent changes of 
the electronic properties of the system. 

Besides the interface region, the 
film surfaces frequently undergo a surface reconstruction.
For the surfaces of transition metal oxides,
this reconstruction can be related to an electron depletion
of $d$-orbitals accompanied by lattice contraction, a process observed on several surfaces like
for instance at Pt(111) \cite{fiorentini}. The charge compensation at the surface can also proceed 
via a formation of oxygen vacancies which bring extra positive charge into the top layer. 
In both cases, the lattice
relaxation and polarization plays a key role and leads to 
changes of chemical bonding and covalencies of surface ions \cite{hamann,ohtomo}.

An example of a film-substrate heterostructure,
which exhibits  a combination of surface and interface reconstruction,
is a system of a LaAlO$_3$ film on top of 
a SrTiO$_3$ substrate. This bilayer attracted much attention recently
due to the discovered transition between metallic and insulating properties \cite{thiel}, possible 
interface magnetism  \cite{brinkman} and superconductivity 
\cite{reyren}.
Electrostatic tuning of the charge carriers allows to switch between superconducting 
and insulating state~\cite{caviglia}. A nanoscale control of the interfacial properties 
has been achieved~\cite{cen} and nanoelectronics has become feasible~\cite{cen09}.

In LaAlO$_3$--SrTiO$_3$-heterostructures, a 
several unit-cell thick LaAlO$_3$ (LAO) film is grown on a mm-thick SrTiO$_3$ (STO) 
substrate. Due to the extremely small thickness of the LAO film, the top
surface states are almost directly coupled to the interface. This coupling leads to a new
cooperative mechanism of 
reconstruction which involves both surface and interface electron states.
As a consequence of such a combination, one may expect a strong dependence of the physical properties on
the thickness of the LAO film. 
In fact, experiments \cite{thiel} with LAO-STO give clear evidence
of an abrupt increase of the carrier concentration and conductivity in heterostructures with
LaO-TiO$_2$ interfaces of $n$-type and AlO$_2$ termination of the LAO film. 
These abrupt changes have a character
of a transition between insulating and metallic states which occurs upon an increase of the 
LAO-thickness above a critical value 
of 3 unit cells in the direction
perpendicular to the interface. As the pure electronic compensation of the interface
polarity would result in 0.5 electrons 
in the interfacial region and in metallic properties, the insulating
character of the system with 
the ultrathin LAO film suggests that other reconstruction mechanisms, 
different from the purely electronic, play a dominant role.

In this work, we present the results of electronic structure calculations of LAO-STO superlattices.
The structure of model superlattices contains the bilayers of LAO and STO with a slab of vacuum
on top of the LAO layer, a structure which closely resembles the experimentally studied LAO-STO heterostructures.
To understand better the surface reconstruction, we compare our results with calculations
for a sandwich-type STO-LAO-STO superlattice
where an LAO unit cell is directly connected to the TiO$_2$-terminated STO layers.
In the analysis of the lattice degrees of freedom, we perform a full structural
relaxation of the considered systems which includes the optimization of the lattice constants and of
the atomic positions in supercells. In our calculations, we employ the SIC variant of 
LSDA+$U$ \cite{wien2k,anisimov} on 
grids containing from 
($8\times 8 \times 1$) to ($9\times 9 \times 1$) $\ve{k}$-points 
where the additional corrections due to Coulomb repulsion on $3d$ orbitals of Ti 
($U_{\rm Ti}=2$~eV) and $5d$ orbitals of La
($U_{\rm La}=8$~eV) are included, according to estimates in Ref.~\onlinecite{bandyopadhyay}.

As appears from our analysis of 
the TiO$_2$-LaO interface stack, the reconstruction in 
STO-LAO-STO sandwiches differs from that in STO-LAO bilayers. The reason is
the significance of the additional distortions in the near-surface planes of LAO in the bilayers, 
whereas in sandwiches we find only small atomic
displacements in the interface LaO layers. 
Moreover, the reconstruction of STO-LAO bilayers depends on the type of
surface termination of the LAO film. Below we consider the consequences of the surface termination.
We will see that, despite similar
TiO$_2$-LaO interface stacking, the structural relaxation in all these systems leads to qualitatively
different properties which demonstrates the key role of the lattice reconstruction.

\section{Interface reconstruction in sandwiches}
\label{chap:II}

\begin{figure}[t]
\epsfxsize=8.5cm {\epsffile{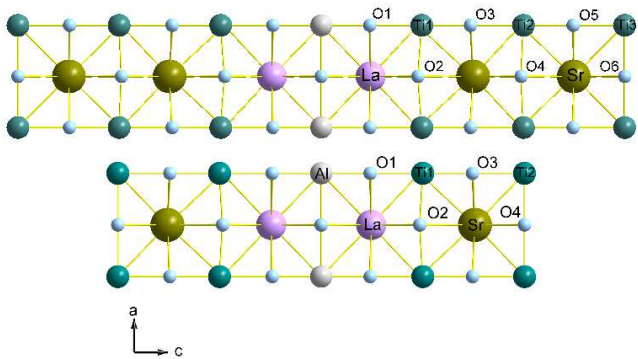}} \caption{
Schematic view of relaxed LAO-STO sandwiches with $N_{\rm LaO}=2$  
planes of LaO and $N_{\rm TiO_2}=2$ and 3 
planes of TiO$_2$ in each substructure. Here, $N_{\rm TiO_2}=2$ 
for the bottom and $N_{\rm TiO_2}=3$ for the top structure, respectively.
The distinct buckling in the interface TiO$_2$ planes 
is evident.} \label{fig1}
\end{figure}

It is well established in the literature~\cite{ohtomo} 
that for the polar (LaO)$^+$--(TiO$_2$)$^0$ interfaces, the 
``polar catastrophe'' can be prevented by an electronic reconstruction mechanism\cite{hesper}.
In sandwiches LAO/STO/LAO/$\ldots$ with LAO terminated by a LaO plane
(see Fig.~\ref{fig1}), electronic reconstruction results in $0.5$ electrons per interface
unit cell required to maintain the overall neutrality of the system. It is remarkable that 
the negative ``extra charge'' is located mostly in the Ti $3d$ 
orbitals.~\cite{gemming,pentcheva} 

In the analysis of heterostructures which contain
SrTiO$_3$ with its incipient ferroelectric properties, one also 
has to take into account the lattice degrees of freedom. 
Due to the high polarizability of 
the titanates, the charge polarity at interface
regions will inevitably induce local atomic distortions. The resulting 
atomic relaxations will modify the interface band structure and metallicity
which can bring radical changes into the electronic reconstruction 
mechanism. 

\begin{figure}[b]
\epsfxsize=8.0cm {\epsffile{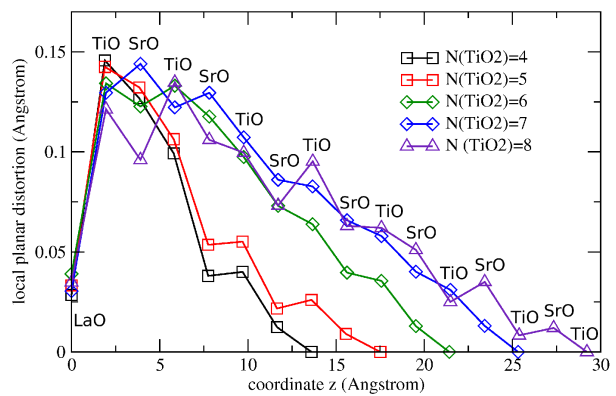}} \caption{
Static dipole distortions in the planes SrO and TiO$_2$ which form with 
increasing number of STO unit cells in a periodic STO-LAO-STO supercell. 
The local planar distortions (vertical axis) are the
changes of the length $\Delta_{AO}=z_A-z_O$ of each (A-O) bond 
(A~=~Ti, Sr, La) with respect to their bulk values. Here the coordinate
$z$ measures the distance from the interface LaO plane of the sandwich LAO/STO/LAO
to each plane in STO.
} \label{fig2}
\end{figure}

In the investigation of a combined electron-lattice reconstruction in LAO/STO,
we first optimized 
the interface distance between TiO$_2$ and LaO layers 
within LDA \cite{wien2k}. 
The optimal distance (1.89~\AA) is shorter then the bond length [Ti-O] (1.953~\AA)
in bulk STO which implies a contraction of the interfacial TiO$_6$ octahedra
in agreement with the results of Ref.~\onlinecite{vonk}.  
After this optimization at the interface, the local positions of atoms 
in the sandwich were relaxed to the new energetically favourable values.
In each plane, the structural relaxation leads to significant, 
diametrically opposed displacements of Ti and O atoms in the 
[001] direction (denoted below by $z$).
As a consequence, 
a static polarization is built up, characterized by the elongation of TiO$_6$ octahedra by about 
0.1~\AA\ in
[001] direction and by a buckling of the TiO$_2$ planes. The
changes of the length $\Delta_{AO}=z_A-z_O$ of each (A-O) bond 
(A~=~Ti, Sr, La) are shown in Fig.~\ref{fig2}. 
These displacements cause a formation of static dipole moments in an initially paraelectric
STO layer
(similarly to what has been found in GGA and GGA+$U$, Ref.~\onlinecite{pentcheva2}) and reflect the degree of buckling 
of LaO, TiO$_2$ and SrO planes. 
A good convergence of the dipole
moments within the interfacial region of $25$~\AA~is achieved already for systems 
containing
seven TiO$_2$-planes in 
the supercell. The dependence of the displacements on the distance from
the interface agrees well with the results of Ref.~\onlinecite{schuster}, which suggest
an exponential decay of the displacements. 
The largest bond distortions 
of 0.15~\AA\ appear in the TiO$_2$ planes and increase for 
increasing thickness of the STO layers. In contrast to the STO layer, the distortions in 
the LAO planes are negligibly small. It is 
worthwhile to note that the most significant atomic displacements
$\Delta \ge 0.05$~\AA\ 
do not extend beyond a thickness of approximately 20~\AA,
which measures an effective depth of structural relaxations near the interface. 

\begin{figure}[t]
\epsfxsize=8.0cm {\epsffile{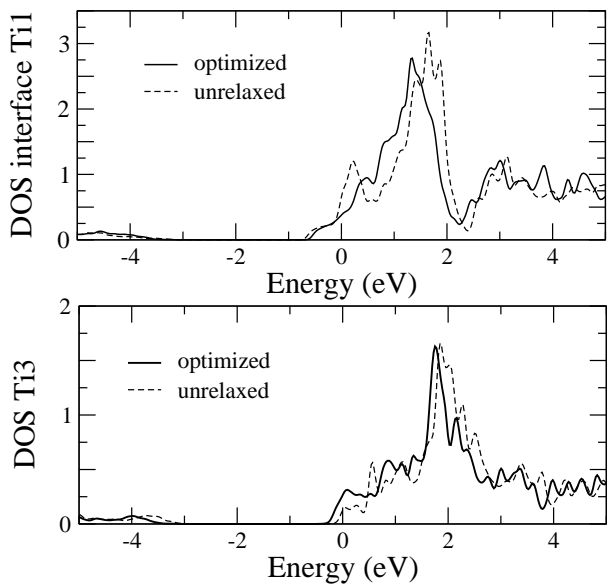}} \caption{
Projected densities of states for Ti1 in the interface TiO$_2$ layer and for Ti3
in the TiO$_2$ plane separated by a distance $2a=7.81$~\AA\ from the interface
in a STO-LAO-STO sandwich with unrelaxed
and optimized atomic positions
($N_{\rm TiO_2}=3 $ and $N_{\rm LaO}=2$).
} \label{fig3}
\end{figure}
  
\begin{figure}[b]
\epsfxsize=8.0cm {\epsffile{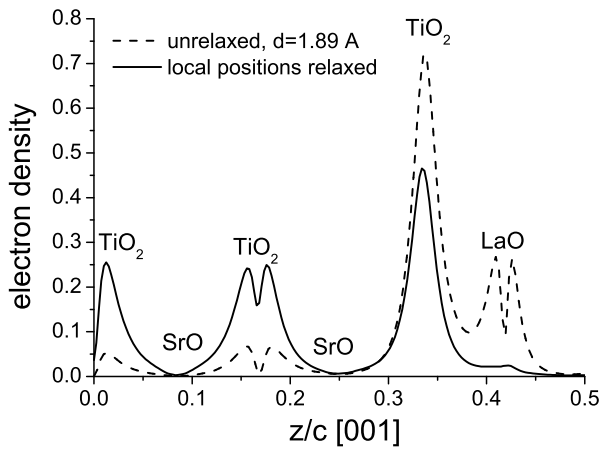}} \caption{
Electron density profiles in structurally unrelaxed and optimized STO-LAO-STO sandwiches
with $N_{\rm TiO_2}=3 $ and $N_{\rm LaO}=2$. The density is determined from the integration
over the density of states of Fig.~\ref{fig3}.
} \label{fig4}
\end{figure} 
  
In the electronic band structure calculated within  LSDA+$U$,
the deformation of TiO$_6$ produces a
Jahn-Teller splitting of t$_{2g}$ states. It should be noted that the t$_{2g}$ 
states are partially filled by the charge carriers appearing due to the electronic
reconstruction of the polar LaO-STO interface. Specifically, the structural
distortions split the t$_{2g}$ levels into $d_{xy}$ and degenerate
$d_{xz}$ and $d_{yz}$ bands. As the atomic displacements strongly depend
on the distance from the interface, the level splitting and related changes
in the band structure will be different in the interface region and in the
layers more distant from the interface. Fig.~\ref{fig3} shows the modifications of
the projected density of states of Ti in the 
layer next to the interface (Ti1) and in 
a more distant (3$^{\rm rd}$)  layer (Ti3). 
In the interface
layer, the relaxation leads to a slight shift of the conduction $3d_{xy}$ band
and, more prominently, to a reduced DOS close to the Fermi energy which
suggests a decrease of the charge carrier density.
In contrast to the interface TiO$_2$, the relaxed $3d_{xy}$ band of 
Ti3 is displaced further below the Fermi level and the DOS at the Fermi energy is enhanced. 
The shifts of the conduction bands also produce changes in their respective occupancies.
They can be calculated through the integration of 
the DOS in the range between the bottom of the
conduction band and the Fermi level.
The calculated profiles of the electron charge density 
(Fig.~\ref{fig4}) suggest a 
charge depletion of the interfacial TiO$_2$-LaO stack and consequent accumulation
of the electron charge in the TiO$_2$ layers which are more distant from the interface.

\begin{figure}[t]
\epsfxsize=8.9cm {\epsffile{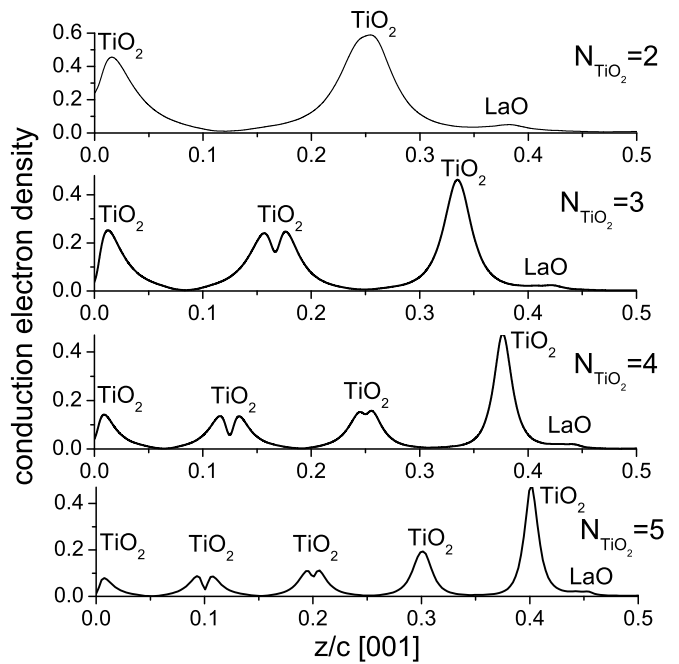}} \caption{
Electron density profiles in structurally optimized STO-LAO-STO sandwiches
with $N_{\rm LAO}=2$ for different $N_{\rm TiO_2}$.
} \label{fig5}
\end{figure}

\begin{figure}[t]
\epsfxsize=8.0cm {\epsffile{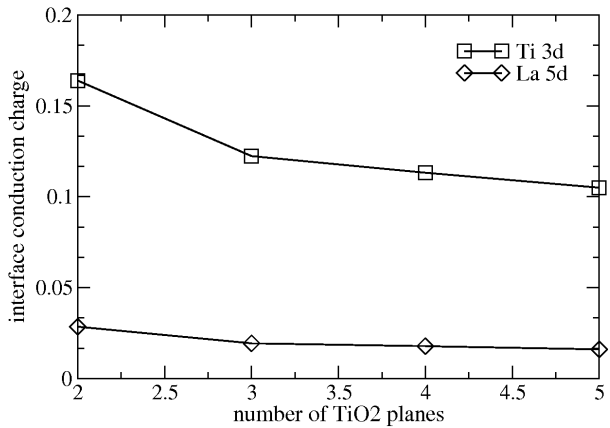}} \caption{
Electron charge as a function of the number of TiO$_2$ planes 
in structurally optimized STO-LAO-STO 
sandwiches with $N_{\rm LaO}=2$.
} \label{fig6}
\end{figure}

The depletion of conducting charge in the interface layers
not only results from the relaxation but may also be triggered through 
a growing number of TiO$_2$ planes in STO. 
This fact can be 
observed in Fig.~\ref{fig5} which demonstrates a redistribution of
a fixed amount (0.5) of compensated charge
with an increasing  number $N_{\rm TiO_2}$ of TiO$_2$ layers from 2 to 5. 
An important consequence of the electronic reconstruction  at the interface 
is a decrease of electron density in the 
TiO$_2$ and LaO 
interface planes (see Fig.~\ref{fig6}) which is stronger
in systems with wider STO substructure. In our calculations, 
an increase of the number of TiO$_2$ planes, $N_{\rm TiO_2}$, 
leads to a decrease of charge in the interfacial TiO$_2$ plane, with a tendency
for a saturation of electron charge in the interface TiO$_2$ plane to a value 
of about 0.1~electrons for a growing number of TiO$_2$ planes in the STO slab.
In the STO layer with 5 TiO$_2$ planes, the
residual $0.4$ electrons are distributed within the more distant 
TiO$_2$ planes, but about $73\%$ of the whole charge is located in the first three 
TiO$_2$ planes. 
It suggests a confinement of electron charge in
an interface region of a thickness of 10--20~\AA\ in the 
STO layer. 
This finding 
agrees well with the experimental distribution
of the Ti excess charge discussed in Ref.~\onlinecite{hamann}, where the distinct
peaks of the excess charge and its substantial decay with 
increasing distance
from the interface demonstrate the confinement of charge to the interfacial 
region of a width of about $30$~\AA. 
Specifically, for sandwiches containing 9 to 11 TiO$_2$ layers,
the Ti excess charge approaches a value slightly above 0.1~electrons at a distance 
of about $3$--$4$~\AA, an experimental value~\cite{hamann} which is 
compatible with the values presented in Fig.~\ref{fig5}.

We reason that the small width of the interfacial region is consistent with the 
conclusion that a quasi-two-dimensional electron gas is formed near the interface
of a LAO-STO heterostructure with a mm-thick STO substrate
(as discussed in Ref.~\onlinecite{thiel}
and, more recently, in Ref.~\onlinecite{sing}).

\section{Role of surface reconstruction for the electronic properties of bilayers}

Experimentally, it has been established that LAO-STO heterostructures with 
more than three unit cells of LAO on STO substrates are metallic. 
For the analysis of the insulator-metal transition with 
increasing thickness of the LAO film, one has
to account for the features inherent to surfaces of transition 
metals and their oxides. The current 
studies of these surfaces are 
motivated by drastic changes of their physical and chemical 
properties with the realization of different types of atomic reconstruction in the surface
and subsurface layers. Furthermore,
a  number of {\it ab-initio} studies of this type of interfaces have identified a metallic 
$n$-type state with charge carriers which predominantly fill
3$\,d_{xy}$ orbitals.\cite{pentcheva,schwingenschloegl}

Recent experiments give indications of
a  temperature dependence of the surface termination
of LAO which changes from AlO$_2$ to LaO planes with an increase of temperature
above 523~K.\cite{vonk,yao,francis} Therefore, to understand the interface reconstruction mechanism in
ultrathin LAO films with an interference between 
surface and interface states,
it is instructive to consider separately two possible types of the surface plane (AlO$_2$
and LaO) in STO-LAO bilayers. 

For our theoretical studies, we have generated supercells which 
contain several LAO unit cells deposited on 1.5 SrTiO$_3$ unit cells. 
We have restricted calculations to 
structures with $N_{\rm TiO_2}= 2$
due to clear indications of charge confinement to a few unit cells of STO found
in the analysis of sandwiches (section \ref{chap:II}).
To consider the surface properties, we have also introduced
a 13~\AA\ vacuum layer on top of LaAlO$_3$. The thickness of the
LaAlO$_3$ film has been varied from 1 to 3 unit cells. In a system with the topmost plane
LaO, an extra half of
a LAO unit cell has been attached to the top of the LAO film. 

The LSDA+$U$ calculations result in a stable nonmagnetic state
of the system. In this state, the electronic properties strongly depend on the structural relaxation
which is controlled by the thickness of the LAO film in the slab. Specifically, in the LSDA+$U$ calculations,
the initially chosen configurations with nonzero magnetic moments and charge inhomogeneities
rapidly decayed to zero magnetic
moments and to equal charge for all Ti atoms in each TiO$_2$ plane which
excludes the possibility for a decisive role of charge or magnetic order
on the resultant electronic state.

\subsection{AlO$_2$ termination}

\begin{figure}[b]
\epsfxsize=6.5cm \epsffile{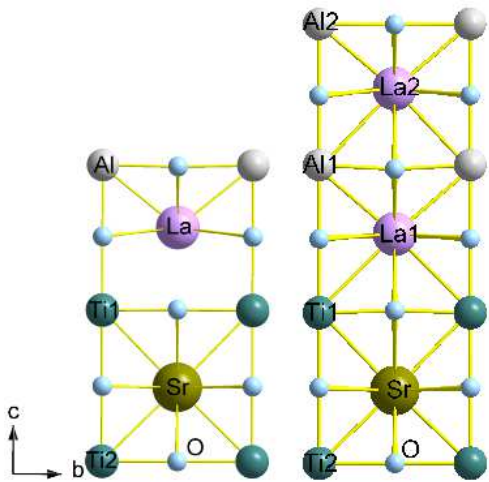} \caption{
Schematic view of relaxed LAO-STO heterostructures with $N_{\rm TiO_2}=2 $ and with top
layers of LaAlO$_3$ terminated by AlO$_2$. 
Here the configurations with 
$N_{\rm LaO}=1$ and $N_{\rm LaO}=2$ are shown. For $N_{\rm LaO}=1$,
the covalencies between La and the titanium and oxygen ions in the plane below
are weak and no bonds are drawn.
%% Das muessen wir doch noch unbedingt angeben, oder?:
The full supercell for our calculations contains, apart
from the vacuum cells, also the inverted structure so that 
the supercell geometry has two inversion symmetric interfaces and surfaces.
} \label{fig7}
\end{figure}

To study the bilayers STO-LAO with
an AlO$_2$ plane at the top surface (Fig.~\ref{fig7}), we have generated a number
of supercells with 
$N_{\rm LaO}$  varying from 1 to 3 LAO unit cells. In these supercells, the
structural optimization results in a structure with large opposite distortions of cations and anions
in the LaO planes and only small atomic displacements in the STO substrate (see Fig.~\ref{fig7}).
Such a relaxation pattern is completely distinct from the systems with LaO termination 
which will be considered in the following section~\ref{lao_subs}. The details of 
the structural changes are presented in 
Table~\ref{tab1}. 

The first important feature is the strong buckling of the LaO planes where
the relative [La-O] displacement
$\Delta z_{\rm LaO} =  3z_{\rm La} - 2z_{\rm O}$ approaches
a maximal value of $0.77$~\AA\ in the structure with 
$N_{\rm LaO}=1$ (cf.\ Fig~\ref{fig7}).
This displacement parameter $\Delta z_{\rm LaO}$ is introduced so as to
characterize the dipolar distortion of a LaO plane.
For $N_{\rm LaO}>1$ the displacement parameter $\Delta z_{\rm LaO}$  is considerably reduced.

\begin{table}
\caption{\label{tab1} Atomic displacements (in \AA) in the top 
planes of STO-LAO bilayers
with an AlO$_2$ plane at the surface. Here the atomic displacements of
the top La sites are defined as $\Delta z_{\rm La}=z_{\rm La}-z^0_{\rm La}$
where $z^0_{\rm La}$ is the planar unrelaxed coordinate, and
of the oxygen sites as $\Delta z_{\rm O}=z_{\rm O}-z^0_{\rm O}$.
An additional index La or Al at $z_{\rm O_{La/Al}}$
refers to the oxygen position in the La or Al planes, respectively.
The atomic distortions 
in the top LaO plane are defined as 
$\Delta z_{\rm LaO}=3\Delta z_{\rm La}-2\Delta z_{\rm O_{La}}$
The buckling of the surface is given by the
atomic distortions in the top AlO$_2$ layer which are defined as
$\Delta z_{\rm AlO_2}=4 \Delta z_{\rm O_{Al}}-3 \Delta z_{\rm Al}$.
\\}

\begin{ruledtabular}
\begin{tabular}{llllllll}
$N_{\rm LaO}$  & $\Delta z_{\rm La}$ & $\Delta z_{\rm
O_{La}}$ &  $\Delta z_{\rm LaO}$ & $\Delta z_{\rm Al}$ & $\Delta z_{\rm
O_{Al}}$ & $\Delta z_{\rm AlO_2}$ \\
\hline
1 & 0.29 & 0.05 & 0.77  & 0.015 & -0.02 & -0.125 \\
2 & 0.11 & -0.07 & 0.46 & -0.11 & -0.13 & -0.19 \\
3 & 0.06 & -0.09 & 0.36 & -0.09 & -0.09 & -0.09
\end{tabular}
\end{ruledtabular}
\end{table}

The second key property is a 
displacement of the topmost AlO$_2$  plane. It comprises 
significant inward displacements of both Al and oxygen towards the LaO plane and
a consequent decrease of the distance between 
the AlO$_2$ plane at the surface and its adjacent LaO plane.

\begin{figure}[t]
\epsfxsize=8.6cm {\epsffile{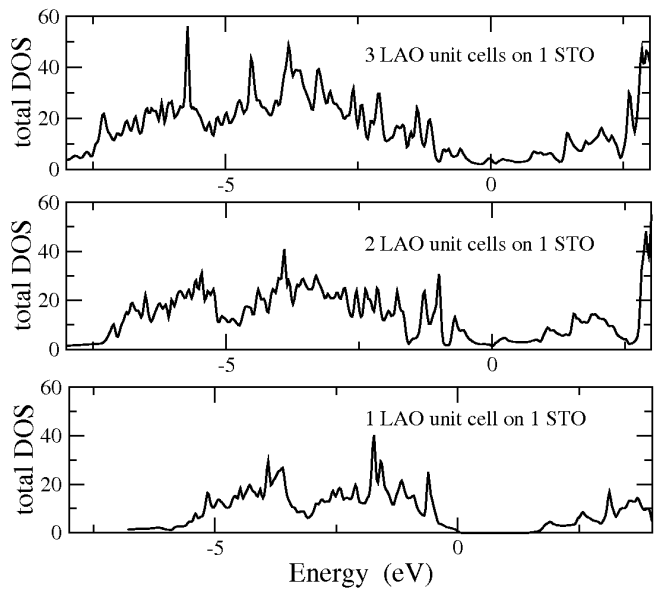}} \caption{
Electron densities of states of a relaxed STO-LAO bilayer
with an AlO$_2$ plane as the surface plane
for $N_{\rm TiO_2}=2$ and different $N_{\rm LaO}$.
} \label{fig8}
\end{figure}

In the case of $N_{\rm LaO}=1$, the electric dipole moment, which is induced by the 
atomic distortions, is given 
by $\Delta P_{\rm LAO} = \sum_i Q_i \Delta z_i $ where $Q_i$ refers to
the charge of the $i$-th site ($i=$ Al, La, O$_{\rm Al}$, O$_{\rm La}$) where O$_{\rm Al/La}$
is the oxygen in the  AlO$_2$/LaO plane. One verifies that
$\Delta P_{\rm LAO} = Z_1\Delta z_{\rm LaO}+Z_2\Delta z_{\rm AlO_2}=0.9$~e\AA\ 
where $Z_1=-Z_2=e$. Here, the displacement parameter for the AlO$_2$ plane is
$\Delta z_{\rm AlO_2}=4 \Delta z_{\rm O_{Al}}-3 \Delta z_{\rm Al}$.

On the other hand,
the polar catastrophe in LAO \cite{nakagawa} leads to the net electric polarization
$P^0_{\rm LAO}=-ea_{\rm LAO}/2=-1.9$~e\AA,
where $a_{\rm LAO}\simeq 3.8$~\AA\ is the experimental bulk 
lattice constant of LaAlO$_3$.
While the interaction of electrons with the polarization $P^0_{\rm LAO}$ tends to close the energy gap
and stabilize the metallic state, the polarization due to atomic distortions is 
of opposite sign with respect to
$P^0_{\rm LAO}$ and leads to a competing tendency to stabilize the insulating state.
In the 
monolayer LAO film with a dielectric constant $\varepsilon$,
the interaction energy
$E=eP_{\rm LAO}/(4\pi\varepsilon_0\varepsilon a_{\rm LAO}^2)\approx -1/\varepsilon$~eV
of the total dipole moment $P_{\rm LAO}=P_{\rm LAO}^0+\Delta P_{\rm LAO}=-1$~e\AA\ 
with the electronic charge 
results in a reduced but finite energy gap of 1.5~eV which,
for comparison, is 
approximately 4~eV in the LDA calculations of bulk LAO.

For the structure with
$N_{\rm LaO}=2$, we have 
$P_{\rm LAO}^0=-N_{\rm LAO} e a_{\rm LAO}/2=-3.8$~e\AA\
and $\Delta P_{\rm LAO}=0.65$~e\AA. 
As a result, the increased net polarization $P_{\rm LAO}=-3.15$~e\AA\ 
closes the energy gap and leads to the metallic state observed
in Fig.~\ref{fig8}.

\begin{figure}[b]
\epsfxsize=8.6cm {\epsffile{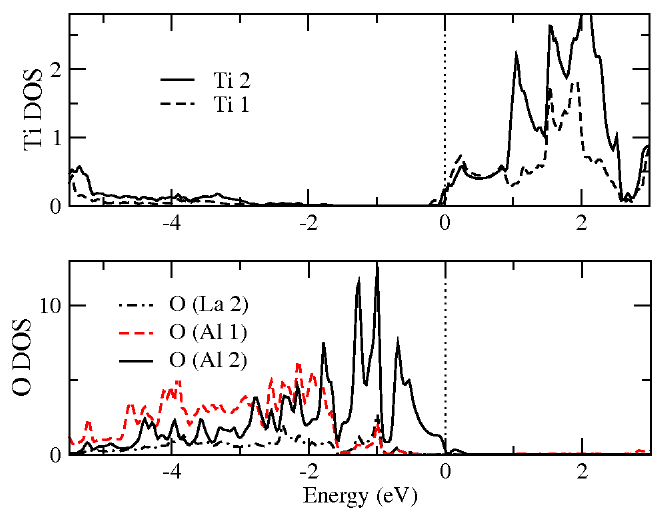}} \caption{
Electron densities of states of Ti and oxygens in the LAO layer of a
relaxed STO-LAO metallic bilayer with $N_{\rm TiO_2}=2$ and 
$N_{\rm LaO}=2$.
} \label{fig9}
\end{figure}

\begin{figure}[t]
\epsfxsize=8.5cm {\epsffile{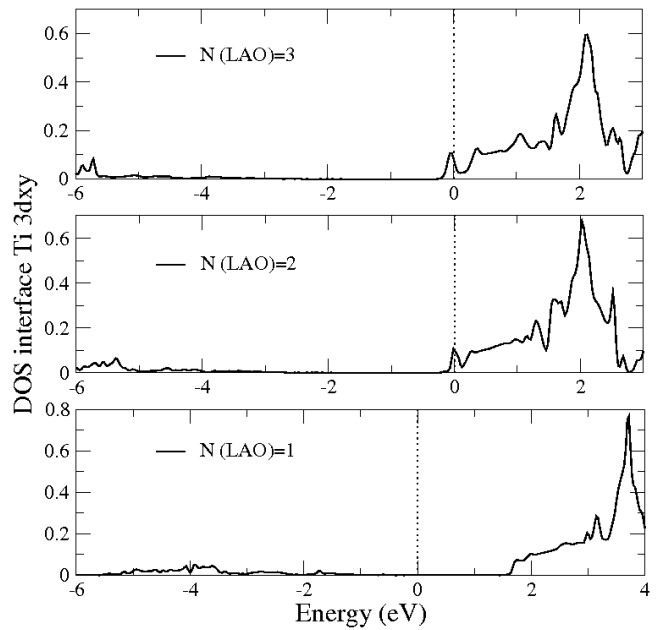}} \caption{
Projected electron densities of states of the interface Ti 3$d_{xy}$ orbitals
for relaxed STO-LAO metallic bilayers with $N_{\rm TiO_2}=2$ and 
$N_{\rm LaO}=1$, 2 and 3 from bottom to top.
} \label{fig10}
\end{figure}

In the densities of states in Fig.~\ref{fig8}, 
the buildup of metallic states proceeds through 
the appearance of the electron carriers in the Ti 3$d_{xy}$ bands 
which is exhibited in Fig.~\ref{fig10}. 
This is accompanied by the filling of the gap which, for 
$N_{\rm LaO}=1$, exists between the oxygen $2p$ and the La $5d$ 
and Ti $3d$ bands.
The $2p$ states of the oxygens in the AlO$_2$ planes approach the Fermi energy from below and the interface Ti 3d$_{xy}$ states 
cross the Fermi energy from above (see Fig.~\ref{fig9}).

In experiments, 
STO-LAO bilayers with LAO films of $N_{\rm LaO}=3$ are still insulating whereas LAO films with $N_{\rm LaO} \geq 4$  have 
been found to be metallic.
In the present LDA-calculations, due
to the underestimation of the energy gap by LDA, the 
``critical thickness'' of the LAO layer
is decreased by 2 and the metallic state occurs already at $N_{\rm LaO}=2$. 
Moreover, other localization effects may play a role in real materials\cite{eckstein,pavlenko_kopp}, 
especially when the density of states at 
the Fermi energy is still small (cf.\ Fig.~\ref{fig9}).
In GGA calculations without lattice relaxation, the transition to a metallic state was identified for 
LAO films with $N_{\rm LaO}=4$ (supplement of Ref.~\onlinecite{cen}). The GGA evaluation with 
lattice relaxation~\cite{pentcheva2008} produces a 
transition at around five monolayers of LAO.

Below the critical LAO-thickness, the polarity compensation via 
self charging (see Chap.~\ref{chap:II}) reduces the energy gap,
a process which however is partially suppressed due to structural distortions in the LAO layer.
Similarly to the GGA results presented in Ref.~\onlinecite{pentcheva2008}, we can conclude here that the
polarization 
in the LAO layer due to structural relaxation competes with the electric field
caused by the polarity catastrophe and in this way establishes the insulating state in
STO-LAO bilayers below the LAO critical thickness. 
In contrast, above the critical thickness the polarity compensation
through the surface and interface charging 
leads to the metallic properties. 

In the considered bilayers, two different charging processes can be 
distinguished which 
are exhibited by the composition of the density of states close to the Fermi energy (see the two
panels of Fig.~\ref{fig9}).
Near the positively charged LaO layer at the TiO$_2$-LaO interface, the compensation produces
the additional negative charge accumulated at Ti $3d_{xy}$ orbitals (see Fig.~\ref{fig9} and Fig.~\ref{fig10}).
In distinction to the interface, the negative polarity near the surface AlO$_2$ plane leads to
additional holes distributed over the $2p$ orbitals of the surface oxygen which is clearly observed
in the bottom part of Fig.~\ref{fig9}. Recently, 
indirect indications for positive surface charge in STO-LAO have been discussed in Ref.~\onlinecite{cen} 
where such a positive surface charge is compensated by oxygen vacancies.
In other experimental conditions, the surface hole charge can be also neutralized
by surface oxidation or by near-surface bonding with negatively charged radicals
which can be a possible reason for 
the insulating state usually found on the 
AlO$_2$  surfaces of LAO-STO bilayers.

\subsection{LaO termination} \label{lao_subs}

The low- and room-temperature experiments are performed with heterostructures,
in which the LAO films are terminated by AlO$_2$ planes in the case
of TiO$_2$-LaO interfaces. There are, however, several experimental indications of a
transformation of the termination plane from AlO$_2$ to LaO, which occurs at temperatures
around 423~K \cite{vonk,yao}. Such
%% "a" eingefuegt:
a transformation is explained through a formation of oxygen vacancies
with simultaneous displacements of Al towards the subsurface layers. Although the
discussed AlO$_2$-LAO transformation remains still an open question, it is highly instructive
to consider LaO terminated LAO films in the bilayers STO-LAO and to compare
the electronic states with the electronic properties of AlO$_2$ terminated layers.

For LaO termination at the topmost surface of the bilayers, we suggest and elaborate 
below that the reconstruction is controlled by the
surface tensile stress which causes a charge occupation of $5d$ orbitals and a
simultaneous contraction of the top surface layers \cite{fiorentini}. 

\begin{figure}[b]
\epsfxsize=6.5cm \epsffile{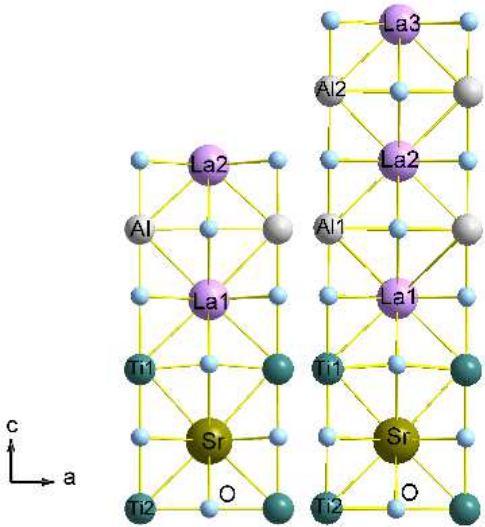} \caption{
Schematic view of
relaxed LAO-STO heterostructures with 
$N_{\rm TiO_2}=2 $ and with 1.5 and 2.5 top unit cells of  LaAlO$_3$ ($N_{\rm LaO}=2$ and 3, respectively).
} \label{fig11}
\end{figure}

In our supercells shown in Fig.~\ref{fig11}, the thickness of the
LaAlO$_3$ film has been varied from 
1.5 to 3.5 unit cells whereas the thickness 
of the SrTiO$_3$ slab was fixed to 
2 TiO$_2$ planes. In the heterostructure,
the polar (LaO)$^{+1}$/TiO$_2$ interface results in an effective electron doping by exactly 
one electron which is needed to achieve  charge neutrality. 
In the context of the recent experiments 
\cite{thiel,brinkman} the key question is 
the influence of the structural relaxation and the concomitant electronic reconstruction on the 
interface state, especially on its metallic and magnetic properties.

\begin{table}
\caption{\label{tab2} Structural parameters (in \AA) for different 
$N_{\rm LaO}$ in STO-LAO bilayers with LaO termination. Here the atomic displacements of 
the top La sites are defined as $\Delta z_{\rm La}=z_{\rm La}-z^0_{\rm La}$
where $z^0_{\rm La}$ is the planar unrelaxed coordinate, and 
of the oxygen sites as $\Delta z_{\rm O}=z_{\rm O}-z^0_{\rm O}$. 
The buckling of the surface is given by
the atomic distortions of the top LaO bond $\Delta z_{\rm LaO}=3z_{\rm La}-2z_{\rm O}$,
and the length of the LaO-bond is denoted by $\Delta_{\rm LaO}$.
The distance between the top LaO and the nearest subsurface plane is $\delta$. 
\\}
\begin{ruledtabular}
\begin{tabular}{llllllll}
$N_{\rm LaO}$ & $a$ & $c$ & $\delta$ & $\Delta z_{\rm La}$ & $\Delta z_{\rm 
O}$ &  $\Delta z_{\rm LaO}$ & $\Delta_{\rm LaO}$ \\
\hline
%% Werte fuer $N_{\rm LaO}$ angepasst:
2 & 3.805 & 32.17 & 1.8 & -0.12 & 0.02 & -0.4 & 2.7 \\
3 & 3.755 & 39.75 & 1.85 & -0.05 & 0.05 & -0.25 & 2.69 \\
4 & 3.759 & 47.32 & 1.85 & -0.05 & 0.05 & -0.25 & 2.66
\end{tabular}
\end{ruledtabular}
\end{table}

\begin{figure}[ht]
\epsfxsize=6.5cm {\epsffile{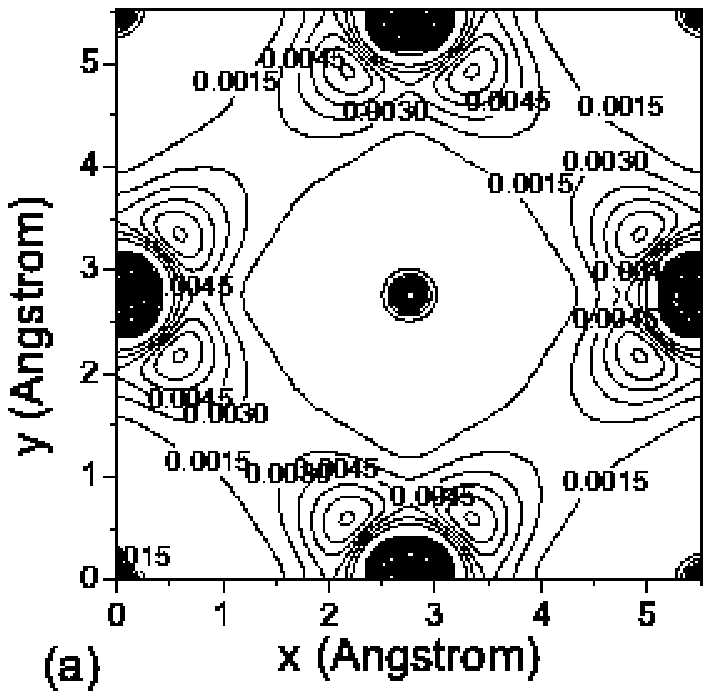}}
\epsfxsize=6.5cm {\epsffile{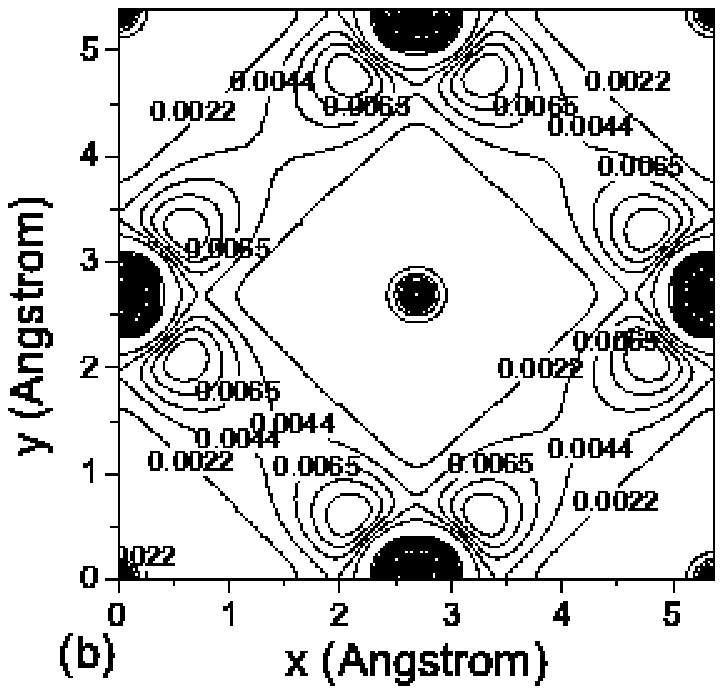}}
\caption{
Electron density map in the ($x$,$y$) plane of (a) unrelaxed and (b) optimized STO-LAO bilayer
with $N_{\rm LaO}=2$ and 
$N_{\rm TiO_2}=2 $. The coordinate $z/c=0.29$ corresponds
to the top surface LaO plane in the LAO film.
} \label{fig12}
\end{figure}

\begin{figure}[th]
\epsfxsize=7.0cm {\epsffile{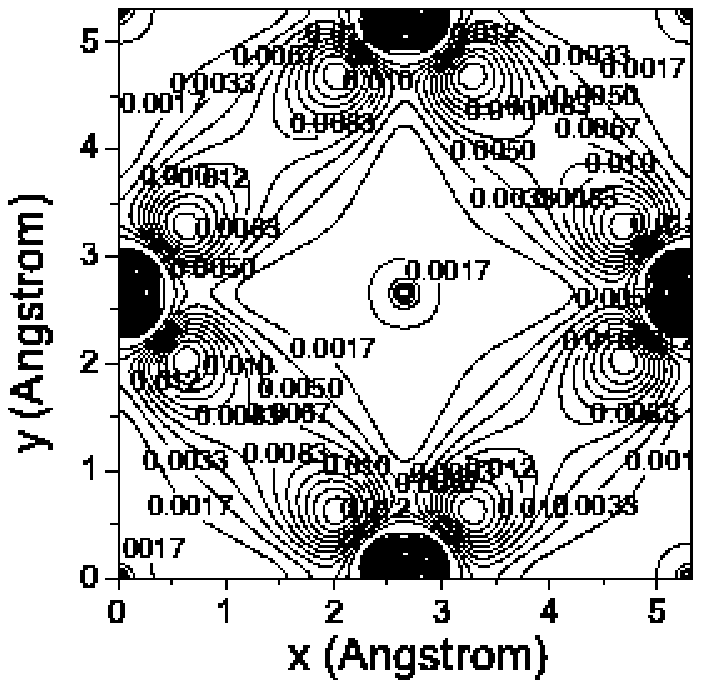}} \caption{
Electron density map in the ($x$,$y$) plane of optimized LAO-STO bilayer
with $N_{\rm LaO}=4$  and 
$N_{\rm TiO_2}=2 $. The coordinate $z/c=0.34$ corresponds
to the top surface LaO plane in the LAO film.
} \label{fig13}
\end{figure}

\begin{figure}[bh]
\epsfxsize=7.0cm {\epsffile{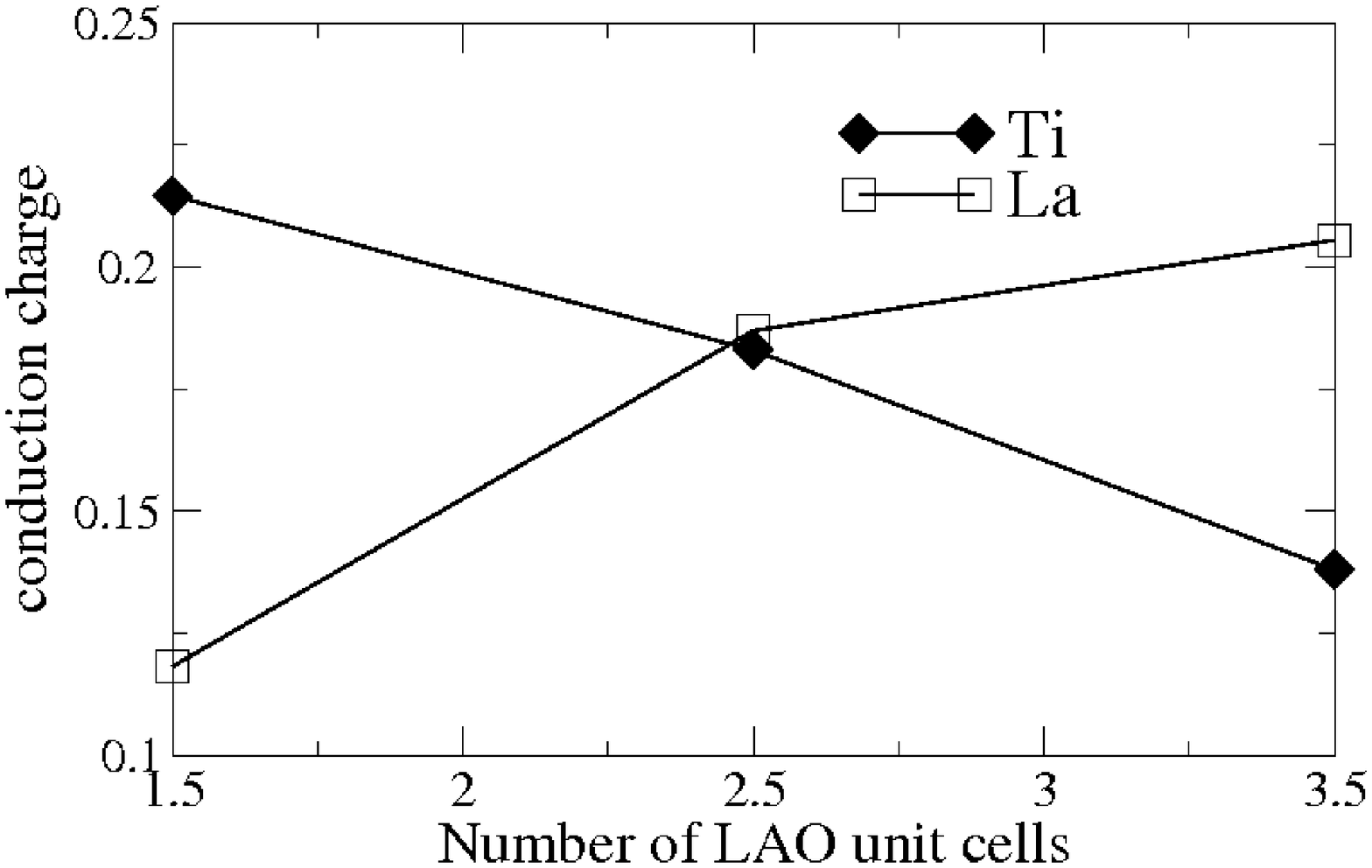}} \caption{
Interface electron charge on La in the top LaO and on Ti in the
interfacial TiO$_2$ planes for different numbers $N_{\rm LaO}$ of LAO planes
in the heterostructure.}
\label{fig14}
\end{figure}

For the optimization of such supercells, we have performed a relaxation of the lattice
constants $a=b$ in the $(x,y)$ plane
and of the subsurface interplanar distances
jointly with a further relaxation of the local atomic
positions in the heterostructure. The results are presented in Table~\ref{tab2}. As appears in
our calculations, the optimized planar lattice constant, $a$, strongly depends on the thickness of the 
LAO film,
varying from 3.805~\AA\ for $N_{\rm LaO}=2$ to 3.76~\AA\ in a supercell with
$N_{\rm LaO}=4$ 
which clearly shows a tendency to approach the bulk lattice constant 
for LaAlO$_3$. It is worth noting that the volume LDA-optimization of bulk LaAlO$_3$ gives
$a^{LDA}_{\rm LAO}=3.75$\AA~ which 
underestimates the experimental value $a_{\rm LAO}=3.789$~\AA.
From Table~\ref{tab2}, one finds that already in ultrathin 
LAO films with $N_{\rm LaO}=3$ and $N_{\rm LaO}=4$ 
the planar lattice parameters $a$ are sufficiently close to the bulk values.

In contrast to the sandwich-type structures characterized by small atomic
displacements in the LAO film, in the considered 
bilayer the atomic relaxation
leads (i) to a substantial change of the distances $\delta$ between the topmost LaO 
and nearest subsurface AlO$_2$ layers and
(ii) to significant opposite displacements of the upper La and oxygen atoms of the top LaO
surface layers. Specifically, we find that
the distance [LaO-AlO$_2$] decreases by a value which ranges from 
0.05~\AA\ in a system with $N_{\rm LaO}=4$ to 0.1~\AA\
for a structure with $N_{\rm LaO}=2$ (see Table~\ref{tab2}).
Furthermore, for larger $N_{\rm LaO}$ the decrease of the La-O dipole moments in $z$ direction 
and of the contraction of the planar lattice constant results in a shortening of length
of the planar [La-O] bond which varies from 
$\Delta_{\rm LaO}=2.7$~\AA\ for $N_{\rm LaO}=2$
to $\Delta_{\rm LaO}=2.66$~\AA\ for $N_{\rm LaO}=4$. 
We also note that in an unrelaxed heterostructure, where the lattice constant
$a$ is fixed to $a_{\rm STO}=3.905$~\AA, the length of the [La-O] bond  
is substantially longer than that of the relaxed system. To apprehend the role of the 
lattice and bond contraction for the electronic 
properties of the subsurface layers, we performed LDA+$U$ 
calculations of the electronic structure of the generated optimized supercells
for different values of $N_{\rm LaO}$.

\begin{figure}[ht]
\epsfxsize=7.0cm {\epsffile{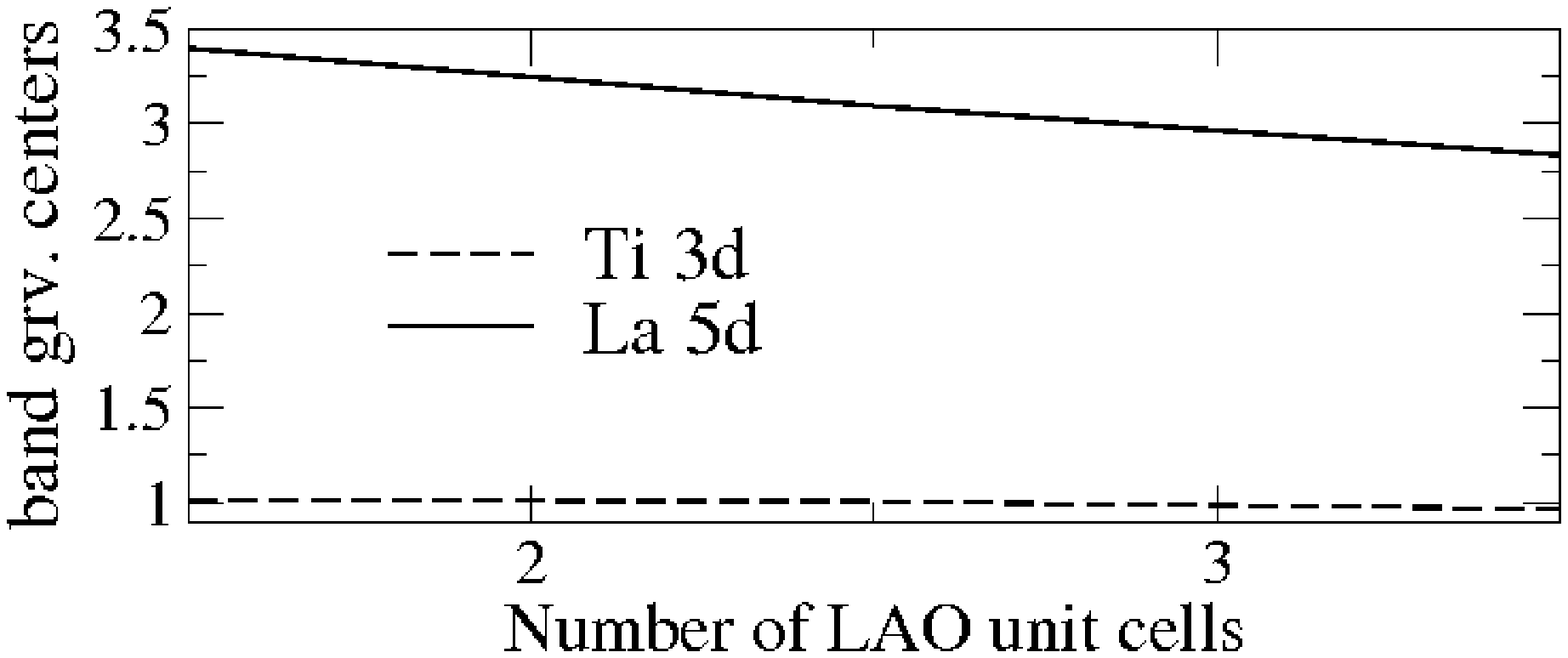}} \caption{
Centers of  gravity of top La $5d_{x^2-y^2}$ and interface Ti $3d_{xy}$ orbitals
versus $N_{\rm LaO}$.}
\label{fig15}
\end{figure}

As compared to the unrelaxed system, in a structure 
with a 1.5 unit-cell LAO film ($N_{\rm LaO}=2$) on SrTiO$_3$,
the Jahn-Teller like splitting due to atomic distortions leads
to an increase of the gap between O $2p$ bands formed by oxygens in TiO$_2$ and LaO planes 
and mixed La $5d$ and Ti $3d$ states, an effect similar to that
in the STO-LAO-STO-sandwiches.
An opposite trend is obtained when we increase the thickness of the LAO film.
Specifically, the thicker LAO layers exhibit smaller
energy gaps and consequently higher charge carrier densities in the top LaO layer.
The structural relaxation plays a central role
for the modifications of the electronic structure.
This is also illustrated in the charge density maps shown in Fig.~\ref{fig12}(a) and
Fig.~\ref{fig12}(b).

\begin{figure}[b]
\epsfxsize=7.5cm {\epsffile{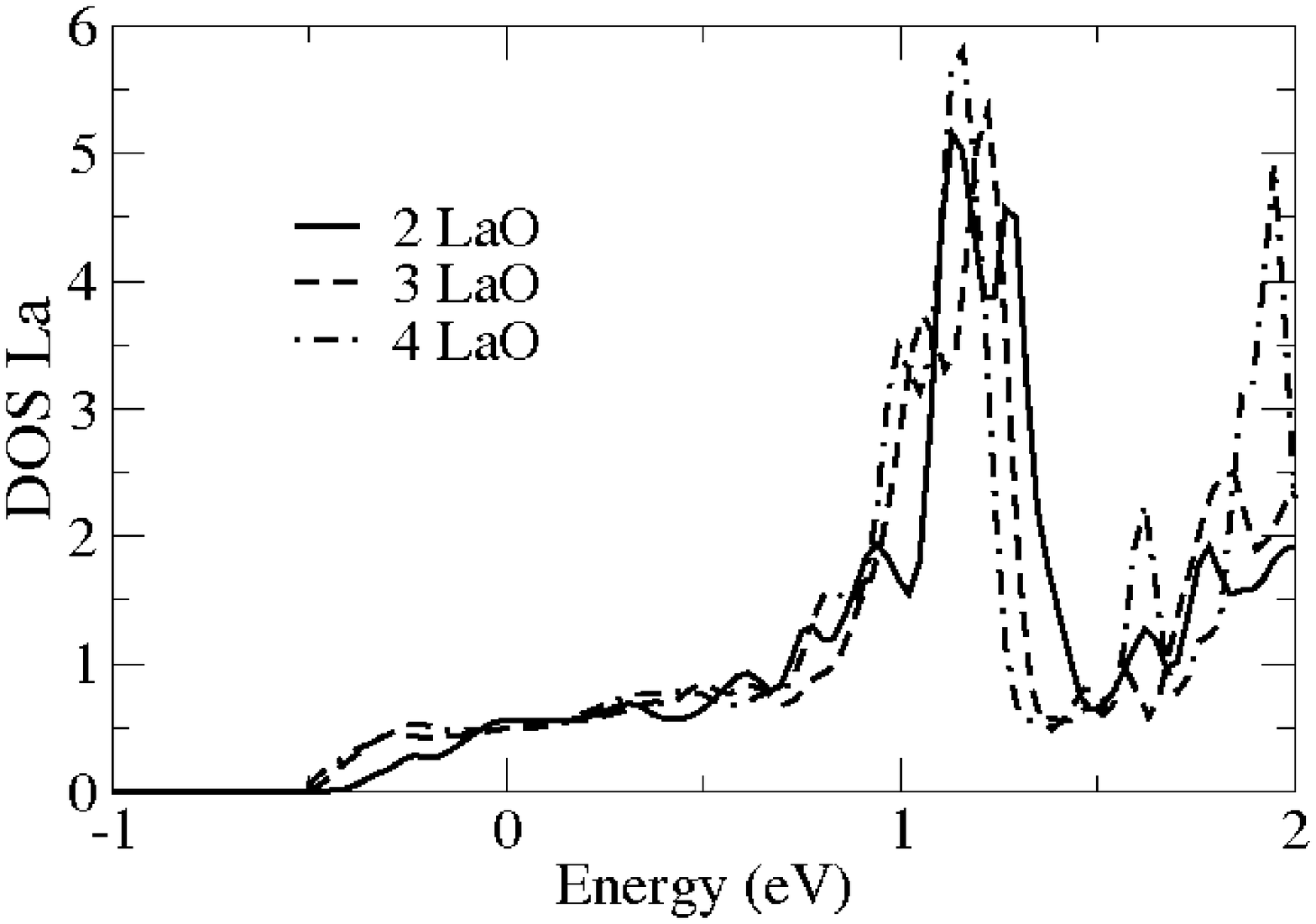}} \caption{
Partial density of states of La in the top LaO plane for different
$N_{\rm LaO}$.}
\label{fig16}
\end{figure}

In the unrelaxed structure with $N_{\rm LaO}=2$,
more charge (about 0.6) is localized on La (Fig.~\ref{fig12}(a)) of the topmost LaO layer.
In the optimized system, the lattice contraction
leads to a strengthening of covalent bonds and to 
a substantial carrier density which is distributed in the space
between La and oxygen ions (Fig.~\ref{fig12}(b)).
In the optimized systems with $N_{\rm LaO} \ge 3$, the increasing contraction
results in a further increase of 
the charge density which is redistributed between 
La and O in the top LaO layer (Fig.~\ref{fig13}).
The obtained feature is closely related to
the changes of the electron occupancy of top La and interfacial Ti 
shown in Fig.~\ref{fig14}. 
As compared to the film with $N_{\rm LaO}=1$, in a thicker LAO film 
with $N_{\rm LaO}=3$ we find a transfer of the carrier density 
from the interfacial TiO$_2$ planes to the top LaO layers
which appears due to the described structural transformations in subsurface layers.
In Fig.~\ref{fig15} this Jahn Teller effect can be clearly detected by the strong 
decrease of the center of gravity of the $5d_{x^2-y^2}$ band of the top La which 
contrasts with the interfacial Ti $3d_{xy}$ band center 
that remains almost unchanged with the variation of the LAO thickness. Here a gravity center 
$E_g$ of a band with a density of states $\rho(E)$ is calculated from the expression 
$E_g={\int dE E \rho(E)}/{\int dE \rho(E)}$. 
The consequential increase of the
electron charge on La 
(Fig.~\ref{fig14}) is also related to the shift of
the bottom of the La $5d_{x^2-y^2}$ band towards lower energies in Fig.~\ref{fig16} and 
is the direct result of the discussed electronic reconstruction.

\section{Conclusions}

We have studied the electronic structure of interfaces between the bulk insulators LaAlO$_3$ and SrTiO$_3$ using 
the local spin density approximation (LSDA) with intra-atomic
Coulomb repulsion (LSDA+$U$). We find that the nature of the interface metallic states
is strongly affected not only by the type (sandwich or bilayer) of the structure
but also prominently by the termination surface of LaAlO$_3$. In all types 
the atomic relaxation plays a crucial role for the electronic properties
of the system. 

In STO-LAO-STO sandwiches the structural relaxation results in an elongation of the
TiO$_6$ octahedra by about 0.1~\AA\ in the  [001] direction and in a buckling of the TiO$_2$ planes
of up to 0.15~\AA\ for the plane nearest to the STO-LAO interface.
These displacements cause a significant polarization in SrTiO$_3$ within about 10~\AA\ from the interface.
They also result in a Jahn-Teller splitting of the  t$_{2g}$ states and the interface charge resides 
mostly in the Ti $3d_{xy}$ bands. The conduction charge is confined essentially 
to the first three TiO$_2$ planes.

In bilayers the surface presents an additional important actor which may control the electronic state
at the interface, especially when the surface is separated by an ultrathin film from the
interface. For a LAO film on top of a STO substrate, the termination (either AlO$_2$ or LaO) at the surface 
is crucial for the atomic relaxation and the electronic state at the LAO-STO interface.

In AlO$_2$-terminated 
bilayers the relaxation occurs primarily in the LaAlO$_3$ film with minor displacements in SrTiO$_3$. 
Specifically, the buckling of the subsurface LaO plane and the modified distances between the 
topmost (oppositely charged) AlO$_2$ and LaO planes introduces a polarization which decreases with
the number $N_{\rm LaO}$ of LAO unit cells.
For the unit cell thick LAO films, it is this lattice relaxation with the corresponding polarization 
of the film which counteracts the polarization
introduced through the self charging (by 1/2 electron to compensate the interface polarity). For films with one unit cell LAO we 
identify an insulating state characterized by an enegy gap of 1.5~eV between the Ti $3d_{xy}$ states of the 
TiO$_2$ plane next to the interface and the oxygen states of the topmost AlO$_2$ plane. For films with 
more than one unit cell LAO (i.e., $N_{\rm LaO}>1$) this gap closes due to the enhanced polarization from the self charging.
Then the metallic state at the interface is formed by the Ti $3d_{xy}$ band
with a finite density of states at the Fermi energy.
This scheme results in a transition from insulating to metallic states:
in the present LDA-calculations, due to the fact that LDA underestimates the gap, 
the metallic state occurs already at $N_{\rm LaO}=2$ whereas experiments show that
bilayers with $N_{\rm LaO}=3$ are still insulating whereas LAO films with $N_{\rm LaO} \geq 4$  have 
been found to be metallic.

Finally, for LaO-terminated bilayers the lattice relaxation is different: the surface tensile stress causes a charge occupation of La $5d$ orbitals and a corresponding contraction of the top surface layers. Here, one electron per interface unit cell is required to  maintain the overall neutrality of the system.  Both, the Ti $3d_{xy}$ states in the plane next to the interface and the La $5d_{x^2-y^2}$have finite density of states at the Fermi energy and stay metallic, irrespective of the number unit cells in the LAO film.

We acknowledge helpful discussions with J. Mannhart and S. Thiel. 
This work was supported through the DFG~SFB-484 
and the TRR~80. 
Grants of computer time from the Leibniz-Rechenzentrum M\"unchen through the HLRB project h1181
are gratefully acknowledged.

\end{document}